\documentclass[12pt]{article}
\usepackage{epstopdf}
\usepackage{epsfig}
\usepackage{graphicx}
\usepackage{a4}
\usepackage{amsmath}
\usepackage{latexsym}
\usepackage{cite}

\usepackage{color}
\usepackage{colordvi}

\graphicspath{{Pics/}}
\DeclareGraphicsExtensions{.eps,.ps}

\usepackage{pslatex}
\usepackage[latin1]{inputenc}

\begin{document}

\begin{titlepage}
\noindent
DESY 13-165 \hfill September 2013\\
LPN 13-060 \\
\vspace{1.3cm}

\begin{center}
  {\bf 
\Large 	
Phenomenology of threshold corrections for inclusive \\[1ex]
jet production at hadron colliders 
  }
  \vspace{1.5cm}

  {\large
    M.C. Kumar$^{\,a}$ 
    and
    S. Moch$^{\,a,b}$
  }\\
  \vspace{1.2cm}

  {\it 
    $^a$ II. Institut f\"ur Theoretische Physik, Universit\"at Hamburg \\
    Luruper Chaussee 149, D--22761 Hamburg, Germany \\
    \vspace{0.2cm}
    $^b$Deutsches Elektronensynchrotron DESY \\
    Platanenallee 6, D--15738 Zeuthen, Germany \\
  }
  \vspace{2.4cm}

\large
{\bf Abstract}
\vspace{-0.2cm}
\end{center}
We study one-jet inclusive hadro-production and compute the QCD threshold corrections 
for large transverse momentum of the jet in the soft-gluon resummation formalism 
at next-to-leading logarithmic accuracy.
We use the resummed result to generate approximate QCD corrections at next-to-next-to leading order, 
compare with results in the literature and present rapidity integrated distributions 
of the jet's transverse momentum for Tevatron and LHC.
For the threshold approximation we investigate its kinematical range of validity 
as well as its dependence on the jet's cone size and kinematics.
\vfill
\end{titlepage}

%
%
\newpage

We study the hadro-production of jets focusing on one-jet inclusive cross sections.
This important scattering process probes parton interactions at very high scales and 
has been measured at the LHC as well as at the Tevatron collider in the
past with very good accuracy \cite{Aad:2011fc, CMS:2011ab, Aaltonen:2008eq, Abazov:2008ae}.
At large momentum transfer the available jet cross section data have not only allowed to set limits in the TeV range 
on the scales of various models for new physics, but have also offered
access to the determination of a number of parameters in Quantum Chromodynamics (QCD).
These include the strong coupling constant $\alpha_s$ as well as the gluon
distribution in the proton at medium to large values of the parton momentum fractions $x$.

In all cases, precise theoretical predictions for the measured rates are an essential prerequisite
and demand good control of the higher order QCD corrections in particular.
It is well-known that these can be sizable and, moreover, are dominated by soft gluon emission 
in the kinematical region where the transverse momentum of the observed jet is large.
At such boundary of phase space the imbalance between virtual corrections and real emission contributions 
gives rise to large logarithms which need to be controlled to high orders in
perturbation theory and, potentially, require resummation.
While the exact next-to-leading order (NLO) results to the $2 \to 2$ parton scattering process 
underlying the one-jet inclusive hadro-production are available since long, 
the computation of the next-to-next-to-leading order (NNLO) cross section predictions 
for $2 \to 2$ parton scattering is yet to be completed. 
In this situation, the threshold logarithms for the one-jet inclusive cross section 
have been used as a means of estimating the size of the exact NNLO QCD corrections~\cite{Kidonakis:2000gi} 
and all-order resummation of soft gluon effects at large transverse momentum
of the identified jet has been achieved~\cite{Kidonakis:1998bk,deFlorian:2005yj,deFlorian:2007fv}.
Recently, the NNLO QCD corrections in the purely gluonic channel 
to one-jet inclusive and di-jet production at hadron colliders has been performed~\cite{Ridder:2013mf}.

In the present paper we perform a phenomenological study of threshold corrections 
to the inclusive jet production at both, Tevatron and LHC 
for the rapidity integrated transverse momentum distributions of the jets. 
To that end, we compute those threshold logarithms 
in the soft-gluon resummation formalism~\cite{Contopanagos:1997nh,Catani:1996yz}
and compare our results at next-to-leading logarithmic (NLL) accuracy with
the available literature~\cite{Kidonakis:2000gi}. 
Given the widespread use of those QCD corrections, e.g., in experimental analysis of 
one-jet inclusive data~\cite{Abazov:2009nc,Abazov:2012lua} 
and in the determination of parton distribution functions (PDFs) from global fits~\cite{Alekhin:2012ig,Alekhin:2012ce,Martin:2009iq}, 
we are particularly interested in assessing the kinematical range of validity of the NLL threshold logarithms. 

For hadro-production of jets the precise definition of the threshold is an important issue, 
because the boundary of phase space for soft gluon emission depends on the details of jet definition, 
i.e., on the jet algorithm, on the jet's cone size and on assumptions of the jet's mass.
As we will see, the resummation of threshold logarithms in~\cite{Kidonakis:2000gi} 
assumes massless jets in the small cone approximation, see~\cite{deFlorian:2007fv}.
In order to scrutinize the threshold approximation, we perform a comparison 
to the exact QCD results at NLO, available e.g., through the programs 
{\tt NLOJET++} \cite{Nagy:2003tz,Nagy:2001fj} or {\tt MEKS}~\cite{Gao:2012he}. 
We find that threshold corrections provide a valid description of the parton dynamics, 
although, within a kinematical range being limited to rather large transverse momenta of jet and to very small jet cone sizes.
Since the latter turn out to be typically much smaller than the currently chosen values at LHC and Tevatron, 
the dependence on finite cone sizes, which is unaccounted for in~\cite{Kidonakis:2000gi}, introduces a large 
additional systematic uncertainty in the threshold approximation.
This is unlike the case of soft-gluon resummation for single-particle inclusive hadro-production 
at high transverse momentum~\cite{Catani:2013vaa,deFlorian:2013taa} 
or for heavy-quark hadro-production (see, e.g.,~\cite{Bonciani:1998vc,Moch:2008qy,Moch:2012mk}), 
where soft-gluon emission is considered relative to a final state composed of on-shell particle(s) 
and the threshold logarithms are found to provide extremely precise predictions through NNLO.

We are considering the following process in proton (anti-)proton collisions at hadron colliders, 
\begin{eqnarray}
P + P( \bar{P}) \rightarrow J + X
\, ,
\end{eqnarray}
where $J$ denotes the observed jet and $X$ the system recoiling against $J$.
At the parton level, a total of $9$ different subprocesses contributes, namely,
\begin{eqnarray}
q(p_1) + q^{\prime}(p_2) &\rightarrow & q(p_3) + q^{\prime}(p_4)\, ,        \nonumber\\
q(p_1) + \bar{q}(p_2)  &\rightarrow & q^{\prime}(p_3) + \bar{q}^{\prime}(p_4)\, , \nonumber\\
q(p_1) + \bar{q}(p_2)  &\rightarrow & q(p_3) + \bar{q}(p_4)\, ,  \nonumber\\
q(p_1) + q (p_2)       &\rightarrow & q(p_3) + q(p_4)\, ,        \nonumber\\
q(p_1) + \bar{q}^{\prime}(p_2) &\rightarrow & q(p_3) + \bar{q}^{\prime}(p_4) \, , \nonumber\\
q(p_1) + \bar{q}(p_2) &\rightarrow & g(p_3) + g(p_4)\, ,        \nonumber\\
q(p_1) + g (p_2)       &\rightarrow & q(p_3) + g(p_4)\, ,        \nonumber\\
g(p_1) + g (p_2)       &\rightarrow & q(p_3) + \bar{q}(p_4)\, ,  \nonumber\\
g(p_1) + g (p_2)       &\rightarrow & g(p_3) + g(p_4)\, . 
\label{eq:2to2}
\end{eqnarray}
The Mandelstam invariants are $s=(p_1+p_2)^2$, $t=(p_1-p_3)^2$ and $u=(p_2-p_3)^2$.
It is to be noted that either of the partons in the final state can give 
rise to the observable jet and the other will be inclusive, implying
that the observable can be computed either by symmetrizing the 
matrix elements between $t$ and $u$ or, alternatively, by 
running the jet-algorithm while doing the phase space integration. 
With these Mandelstam invariants, the relation $s_4 = s+t+u \ge 0$ holds 
where $s_4$ is the invariant mass of the system recoiling against the 
observed jet and $s_4=0$ at threshold.

The perturbative expansion of the partonic cross section $\hat{\sigma}$ 
in powers of the strong coupling constant $\alpha_s$ reads
\begin{eqnarray}
\hat{\sigma} \,=\, \sum\limits_{l = 0}^{\infty} \hat{\sigma}^{(l)} \, ,
\end{eqnarray}
where $\hat{\sigma}^{(0)}$ denotes the Born term.
At higher orders the parton cross section $\hat{\sigma}^{(l)}$ contains plus-distributions 
of the type $\alpha_s^l\,[\ln^{2l-1}(s_4/p_T^2)/s_4]_+$ that lead to the Sudakov logarithms upon 
integration.  In a physical interpretation $s_4$ denotes the additional energy carried away
by real emission of soft gluons above the partonic threshold. 

The generic $l$-loop expanded resummed results can be written as
\begin{eqnarray}
\label{eq:sigmaexp}
\frac{d^2 \hat{\sigma}^{(l)}}{dt~du} & = & 
\sum\limits_{k=0}^{2l-1}\, C_{l,k} \left[ \frac{\hbox{ln}^{(2l-1)-k}\left(s_4/p_T^2\right)} {s_4} \right]_{+} 
+
C_{l,\delta} \delta\left(s_4\right)
+
{\cal O} \left(s_4\right)
\, ,
\end{eqnarray}
and at each loop order, the coefficients $C_{l,0}$ determine the leading logarithm (LL), 
the coefficients $C_{l,1}$ determine the NLL contributions and so on.
It is well-established, that the threshold logarithms exponentiate 
and at the differential level (one-particle inclusive kinematics~\cite{Laenen:1998qw})
this exponentiation has been performed to NLL accuracy in~\cite{Kidonakis:2000gi},  
where the resummed result has been used to generate the results 
in fixed-order perturbation theory through NNLO.

The resummation is based on the factorization of the partonic cross section near threshold into various functions, 
each of which organizes the large corrections stemming from a particular region of phase space. 
The full dynamics of collinear gluon emission from initial or final state partons are summarized in
so-called jet functions ${\cal J}^I$ and ${\cal J}^F$ which contain all LL and some NLL enhancements. 
Additional soft gluon dynamics at NLL accuracy which are not collinear to one
of the external partons are summarized by the soft function $S$, 
which is governed by anomalous dimension $\Gamma_S$~\cite{Contopanagos:1997nh,Kidonakis:1998bk}.
Finally, the effects of off-shell partons are collected in a so-called hard
function $H$, where both $H$ and $S$ are matrices in the space of color
configurations for the respective underlying $2 \to 2$ scattering process in Eq.~(\ref{eq:2to2}).

The resummation is conveniently carried out in the space of moments $N$. 
The formal definition of Laplace moments as 
\begin{eqnarray}
\label{eq:laplace}
{\tilde f}(N) = \int \frac{ds_4}{s}\,e^{-N s_4/s} f(s_4/s) 
\, ,
\end{eqnarray}
establishes the correspondence between the plus-distributions 
for $s_4 \to 0$ and the moments $N \to \infty$, that is $[\ln^{2l-1}(s_4/p_T^2)/s_4]_+ \leftrightarrow \ln^{2l} N$, 
see, e.g.,~\cite{Laenen:1998kp} for details.
Thus, the parton level resummed cross section for a generic subprocess in Eq.~(\ref{eq:2to2}) 
is given by~\cite{Laenen:1998qw,Kidonakis:2000gi}
\begin{eqnarray}
  \nonumber 
  d\hat{\sigma}_{12 ~\to ~34}^{\mbox{\small res}}(N) 
  &=& 
  \mbox{exp} \left[ -\sum_{a=1,2} 2 \int\limits_{\mu_F}^{2p_a.\zeta} \frac{d\mu}{\mu} C_{(f_a)} \frac{\alpha_s (\mu^2)}{\pi} \mbox{ln}N_a\right]
  \\
  \nonumber {}&& 
  \times 
  \mbox{exp} \left[\sum_{a=1,2} {\cal J}^I_a (N_a) \right] \times  \mbox{exp} \left[\sum_{b=3,4} {\cal J}^F_b (N) \right] 
  \\
  \nonumber {} && 
  \times~
  \mbox{exp} \left[2 \sum_{a=1,2} \int\limits_{\mu_F}^{p_T} \frac{d\mu}{\mu}\gamma_a[\alpha_s(\mu^2)]\right] \times
  \mbox{exp} \left[ 4\int\limits_{\mu_R}^{p_T} \frac{d\mu}{\mu} ~ \beta(\alpha_s(\mu^2))\right] \\
  \nonumber {} && 
  \times 
  \mbox{Tr}\Bigg\{ H(\alpha_s(\mu_R^2) ) ~ \bar{P} \mbox{ exp}\left[ \int\limits_{p_T}^{p_T/N} \frac{d\mu}{\mu} ~ \Gamma_S^{\dagger}(\alpha_s(\mu^2))\right] 
  \\
  {} && \times~
  S(\alpha_s(p_T^2/N^2)) ~P \mbox{ exp}\left[\int\limits_{p_T}^{p_T/N} \frac{d\mu}{\mu} ~ \Gamma_S (\alpha_s(\mu^2))\right]
  \Bigg\}
  \, ,
  \label{eq:sigmares}
\end{eqnarray}
where the trace operation acts on the matrices $S$, $H$ and $\Gamma_S$ in color space
and $P$, $\bar P$ denote (complex) ordered matrix products.
The function $\beta$ is the standard QCD beta function, 
$\gamma_q = (\alpha_s / \pi) (3C_F/4)$ and $\gamma_g = (\alpha_s/ \pi)(\beta_0/4)$ are
the anomalous dimensions for quarks and gluons needed to 1-loop accuracy here.
$C_{(f_a)}$ is the quadratic Casimir operator with $C_f = C_F = (N_c^2-1)/(2N_c)$ 
for an external quark/antiquark and $C_f = C_A = N_c$ for an external gluon
with $N_c$ being the number of colors.
The renormalization and factorization scale are given by $\mu_R$ and $\mu_F$.
Moreover, $\zeta_\mu$ is a dimensionless vector specifying the kinematics, 
see~\cite{Laenen:1998qw}, so that in single-particle inclusive kinematics
it can be taken as $\zeta_\mu = p_J/p_T$ and, likewise, the moments $N_a$ ($a=1,2$) 
are given by $N_1 = N(-u/s)$ and $N_2 = N(-t/s)$.

The initial state functions ${\cal J}^I_a$ 
generate the LL and some NLL logarithms as a double integral over the cusp
anomalous dimension $A^{(f_a)}(\alpha_s) = C_f \left((\alpha_s/\pi) + (K/2) (\alpha_s/\pi)^2 \right)$ 
with $K = C_A(67/18-\pi^2/6) -5n_f/9$ and $n_f$ being the number of quark flavors. 
In Mellin space, the ${\cal J}^I_a$ are given by
\begin{eqnarray}
{\cal J}^I_a (N_a) =
-\int_0^1 dz \frac{z^{N_a-1}}{1-z} \left[\int_{(1-z)^2}^{1}
\frac{d\lambda}{\lambda} A^{(f_a)} [\alpha_s(\lambda(2p_a.\zeta)^2)]
\right.
\left.
+\frac{1}{2} \nu^{(f_a)} [\alpha_s((1-z)^2(2p_a.\zeta)^2 )]\right],
\end{eqnarray}
where $\nu^{(f_a)} = 2C_{(f_a)} (\alpha_s/\pi)$.

The final state jet functions ${\cal J}^F_b$ describe both, soft and
hard, radiation collinear to the outgoing partons 
giving rise to the observed jet and the inclusive remainder
recoiling against the observed jet.  
The ${\cal J}^F_b$ are given by
\begin{eqnarray}
\nonumber
{\cal J}^F_b (N) 
&=& 
\int_0^1 dz \frac{z^{N-1}}{1-z} \Bigg[\int_{(1-z)^2}^{(1-z)}
\frac{d\lambda}{\lambda} A^{(f_b)} [\alpha_s(\lambda p_T^2)]
\\
&&
\qquad\qquad
 + B^{(1)}_b[\alpha_s((1-z)p_T^2)] + B^{(2)}_b[\alpha_s((1-z)^2p_T^2)] \Bigg],
\end{eqnarray}
where $B^{(1)}_{(q)} = (-3C_F/4) (\alpha_s/\pi)$, $B^{(2)}_{(q)} = C_F [\mbox{ln}(2\nu_q)-1] (\alpha_s/\pi)$, 
$B^{(1)}_{(g)} = (-\beta_0/4) (\alpha_s/\pi)$ and $B^{(2)}_{(g)} = C_A [\mbox{ln}(2\nu_g)-1] (\alpha_s/\pi)$,
with $\beta_0$ being the first coefficient of the QCD beta function.
Here, the $\nu_i = (\beta_i.n)^2/|n|^2$ are gauge dependent terms, where
$\beta_i = p_i \sqrt{2/s}$ are the particle velocities and $n$ is the axial
gauge vector chosen such that $p_i \cdot \zeta = p_i \cdot n$.
As we have discussed already above, it is in the expression for ${\cal J}^F_b$, that any 
dependence on the jet definition, in particular on the jet's cone size $R$ is lacking. 
This has important consequences, as any finite $R$ dependence will alter the
resummed cross section at LL accuracy, 
since the large logarithms generated by the collinear contributions in ${\cal J}^F_b$ 
are actually regularized by the cone size and 
instead give rise to logarithmic terms in $R$ in the perturbative cross section, 
see also~\cite{deFlorian:2007fv}.
Thus, Eq.~(\ref{eq:sigmares}) holds in the limit $R \to 0$ 
and the numerical impact of such approximation will be illustrated in what follows
when comparing to NLO results for $R$ values typically used in jet analysis.

To investigate this further requires considering the differences between the threshold corrections
and the fixed order results 
by going into the details of their computation, in particular the jet algorithm being used in the NLO computation. 
The higher order QCD corrections crucially depend on the value of 
the parameter $R$ (cone size) used in the jet algorithm. 
A parton in the final state resulting from a hard scattering is completely different from a jet that is observed in the experiments.
At LO the transverse momenta of the two partons in the final state, which eventually hadronize and form 
two jets, balance each other and are well separated in the rapidity-azimuthal angular plane. 
Hence the LO theory predictions are insensitive to the value of $R$.  
However, at NLO and beyond there are additional partons in the final state. 
Whenever two or more partons fall within a cone of size $R$, their momenta are combined 
in a scheme to form a new object which eventually hadronizes to form a single jet. 
The larger the value of $R$, the larger will be the number of jet events thus counted. 
Thus, the higher order QCD corrections for inclusive jet production depend on the value of $R$ and, 
in fact, increase with $R$.
The computation of the threshold corrections on the other hand is based on the phase space slicing underlying 
Eq.~(\ref{eq:sigmares}) and involves the $s_4$ integration which captures the 
information of the additional gluon radiation at higher orders. 
However, there is no explicit additional gluon radiation in the final state that can be subjected to a jet algorithm 
and can eventually be associated with a parton inside a cone of size $R$ to form a single jet.
Thus the threshold corrections Eq.~(\ref{eq:sigmares}) carry no dependence on $R$.

Finally, the soft and the hard functions carry the information about the 
color exchange in the specific parton scattering process 
and account for the associated soft gluon effects in QCD hard scattering.
In our analytical computation we use Symbolic Manipulation program {\tt FORM} 
\cite{Vermaseren:2000nd} and the related color package \cite{vanRitbergen:1998pn} 
for color algebra. Following~\cite{Kidonakis:1998nf} we choose for a $qq \to qq $ 
process $ij \to kl$ the $t$-channel color basis
\begin{eqnarray}
\label{eq:qqbasis}
c_1 = \delta_{ik}~\delta_{jl}\, , \qquad 
c_2 = t^c_{ki}~t^c_{jl}\, ,
\end{eqnarray}
where $t^c_{ij}$ are the generators of $SU(3)$ group in the fundamental representation and 
$N_c=3$ is the number of colors, 
so that the tree level soft function for this basis 
given by $S^{(0)}_{qq \to qq} = {\rm diag}(9, 2)$.
Likewise, the $t$-channel color bases for the $qg \to qg$ process $ij \to kl$ are given by 
\begin{eqnarray}
\label{eq:qgbasis}
c_1 = \delta_{ik} \delta_{jl}\, , \qquad 
c_2 = d^{jlc}t^c_{ki}\, , \qquad \text{and} \qquad 
c_3 = i~f^{jlc}~t^c_{ki}\, ,
\end{eqnarray}
with the tree level soft function 
$S^{(0)}_{qg \to qg} = {\rm diag}(24, 20/3, 12)$
and for a $gg \to gg$ process $ij \to kl$ by 
\begin{eqnarray}
\label{eq:ggbasis}
c_{1,2} &=& \frac{i}{4}[f^{ijm}d^{klm}~ \mp~ d^{ijm}f^{klm}] \, , \nonumber \\
c_3 &=& \frac{i}{4}[f^{ikm}d^{jlm}~ +~ d^{ikm}f^{jlm}] \, ,\nonumber \\ 
c_4 &=& \frac{1}{8} \delta_{ik}\delta_{jl} \, ,\nonumber \\
c_5 &=& \frac{3}{5} d^{ikn}~d^{jln} \, ,\nonumber \\
c_6 &=& \frac{1}{3} f^{ikn}~f^{jln} \, ,\nonumber \\
c_7 &=& \frac{1}{2}\left(\delta_{ij}\delta_{kl} -\delta_{il}\delta_{jk} \right) 
 -\frac{1}{3} f^{ikn}~f^{jln} \, , \nonumber \\
c_8 &=& \frac{1}{2}\left(\delta_{ij}\delta_{kl} +\delta_{il}\delta_{jk} \right)
-\frac{1}{8} \delta_{ik}\delta_{jl}  -\frac{3}{5} d^{ikn}~d^{jln} \, .
\end{eqnarray}
In the latter case, the soft function assumes the form 
$S^{(0)}_{gg \to gg} = {\rm diag}(5, 5, 5, 1, 8, 8, 20, 27)$ for this basis.
All other $2 \to 2$ processes in Eq.~(\ref{eq:2to2}) are obtained by crossing
and together with the corresponding hard functions $H^{(0)}_{ij \to kl}$
the trace ${\rm Tr}(H^{(0)} S^{(0)})$ is proportional to the Born cross section.

The resummation of the soft color exchange requires the computation of the 
soft anomalous dimensions~\cite{Contopanagos:1997nh}, 
where the 1-loop expression $\Gamma_S^{(1)}$ suffices to NLL accuracy.
The soft anomalous dimension is gauge dependent and to 1-loop level it
can be expressed in color space as
\begin{eqnarray}
\Gamma_{S,~IJ} = \Gamma_{S,~IJ}^{(1)} + \delta_{IJ} ~ \frac{\alpha_s}{\pi}
\sum_{i=1}^4 C_{(f_i)} \frac{1}{2} \left[-\mbox{ln}(2\nu_i) +1 - i \pi \right]
\, ,
\end{eqnarray}
where the gauge dependent terms $\nu_i$ are as defined previously.
For the process $qq \to qq$ and in the basis Eq.~(\ref{eq:qqbasis}) 
it is given by
\begin{eqnarray}
\label{eq:gamqq}
\Gamma_{S,\,qq \rightarrow qq}^{(1)}
&=&
\frac{\alpha_s}{\pi}
\begin{bmatrix}
-\frac{1}{3}(T+U)+\frac{8}{3} U & 2U \\
\frac{4}{9}U & \frac{8}{3}~T
\end{bmatrix}
\, ,
\end{eqnarray}
where $T = \hbox{ln}\left(\frac{-t}{s}\right) + i \pi$ and $U = \hbox{ln}\left(\frac{-u}{s}\right) + i\pi$. 
Likewise, for the $qg \to qg$ process in the basis Eq.~(\ref{eq:qgbasis}) we have 
\begin{eqnarray}
\label{eq:gamqg}
\Gamma_{S,\, qg \rightarrow qg}^{(1)} \quad = \quad
\frac{\alpha_s}{\pi}
\begin{bmatrix}
\frac{13}{3}~T & 0 & U \\
0 & \frac{4}{3} T+\frac{3}{2}U & \frac{3}{2}U \\
2U & \frac{5}{6}U  & \frac{4}{3} T + \frac{3}{2}U
\end{bmatrix}
\, ,
\end{eqnarray}
and for the subprocess $gg \rightarrow gg$, cf. Eq.~(\ref{eq:ggbasis}), 
the block-diagonal form
$\Gamma_{S,\, gg \rightarrow gg}^{(1)} = {\rm diag}(G_{3 \times 3},G_{5\times5})$ 
where 
$G_{3\times 3} = (\alpha_s/\pi) {\rm diag}(3T,3U,3(T+U))$
and
\begin{eqnarray}
\label{eq:gamgg}
G_{5\times 5} = \frac{\alpha_s}{\pi}
\begin{bmatrix}
6T & 0 & -6U & 0 & 0 \\
0 & 3T+\frac{3}{2}U & -\frac{3}{2}U & -3U & 0 \\
-\frac{3}{4}U & -\frac{3}{2}U & 3T+\frac{3}{2}U & 0 & -\frac{9}{4}U  \\
0 & -\frac{6}{5}U & 0 & 3U & -\frac{9}{5}U  \\
0 & 0 & -\frac{2}{3}U & -\frac{4}{3}U & -2T+4U
\end{bmatrix}
\, .
\end{eqnarray}

Within this set-up we have computed the resummed cross section in Eq.~(\ref{eq:sigmares}) for all parton channels 
and expand the resummed results to 2-loop level at NLL accuracy.  
At the $1$-loop level, this determines the coefficients 
$C_{1,0}$ and $C_{1,1}$ in Eq.~(\ref{eq:sigmaexp}), while 
the coefficient $C_{1,\delta}$ of the $\delta(s_4)$ 
includes the 1-loop corrections to the hard and the soft function,
$H^{(1)}$ and $S^{(1)}$ that can be extracted from the finite parts 
of the fixed order NLO computation. 
This matching is required for next-to-next-to-leading logarithmic (NNLL) 
contributions and the necessary formulae in various kinematics 
have been derived in~\cite{Kelley:2010fn,Catani:2013vaa}.
At the $2$-loop level Eq.~(\ref{eq:sigmares}) determines $C_{2,0}$ and $C_{2,1}$. 
Starting from NNLL accuracy the coefficient $C_{2,2}$ involves the
hard matching functions mentioned above, i.e., the term $C_{1,\delta}$. 
In the present analysis, though, we have not included these matching functions 
and leave them for future study.

We find that our analytical results for all parton level cross sections are 
in good agreement with those given in~\cite{Kidonakis:2000gi} except 
for a small difference of an overall color factor of $[N_c^2 / (N_c^2-1)^2]$
at NLL level for the subprocess $gg \to q\bar{q}$.
The $1$-loop corrections to NLL accuracy for this subprocess are
\begin{eqnarray}
\nonumber
s^2 \frac{d^2 \hat{\sigma}_{gg \to q\bar{q}}^{(1)}}{dt~du} & = &
\alpha_s 
\hat{\sigma}_{gg \to q\bar{q}}^{(0)} 
\Bigg\{ 
(4 C_A-2 C_F)  \left[ \frac{\hbox{ln}\left(s_4/p_T^2\right)} {s_4} \right]_{+} \\
\nonumber
{} && + \left[
  - 2 C_A \hbox{ln}\left(\frac{\mu_F^2}{p_T^2} \right) 
  - (2 C_F - C_A) \text{ln} \left(\frac{p_T^2}{s}\right)  
  - \frac{3}{2} C_F \right]
\left[\frac{1}{s_4}\right]_{+} \Bigg\} \\
\nonumber
{} &&
+\alpha_s^3
{
\frac{N_c^2}{(N_c^2-1)^2}
}
\Bigg\{
-\frac{(N_c^2-1)}{2N_c^2} \frac{(t^2+u^2)}{tu} \hbox{ln}\left(\frac{p_T^2}{s}\right) \\
{} &&
-\frac{(N_c^2-1)}{2}\left[\frac{u^2-t^2}{tu} + \frac{2(u-t)}{s} \right]\hbox{ln} 
\left(\frac{u}{t} \right) \Bigg\}\left[\frac{1}{s_4}\right]_{+}
\, , 
\label{eq:gg2qqb1l}
\end{eqnarray}
where 
$\hat{\sigma}_{gg \to q\bar{q}}^{(0)}$
contains the spin and color averaged leading order (LO) matrix elements and is given by
\begin{eqnarray}
\hat{\sigma}_{gg \to q\bar{q}}^{(0)} = \alpha_s^2 \left[\frac{1}{6}~\frac{t^2+u^2}{tu} 
- \frac{3}{8}\frac{t^2+u^2}{s^2}\right]
\, .
\end{eqnarray}
The corresponding $2$-loop corrections at NLL accuracy are given by
\begin{eqnarray}
\nonumber
s^2 \frac{d^2 \hat{\sigma}_{gg \to q\bar{q}}^{(2)}}{dt~du} & = &
\left(\frac{\alpha_s^2}{\pi} \right)
\hat{\sigma}_{gg \to q\bar{q}}^{(0)}
\Bigg\{ 
\frac{1}{2}(4 C_A-2 C_F)^2  \left[ \frac{\hbox{ln}^3\left(s_4/p_T^2\right)} {s_4} \right]_{+} \\
\nonumber
{} && + \Bigg[ 3(2 C_A - C_F)\left[ 
  - 2 C_A \hbox{ln}\left(\frac{\mu_F^2}{p_T^2} \right) 
  - (2 C_F - C_A) \text{ln} \left(\frac{p_T^2}{s}\right) 
  - \frac{3}{2} C_F \right] \\
\nonumber
{} && + \beta_0 \left(-C_A + \frac{3}{4} C_F \right) \Bigg]
\left[\frac{\hbox{ln}^2(s_4/p_T^2)}{s_4} \right]_{+} \Bigg\} \\
\nonumber
{} && 
+\frac{\alpha_s^4}{\pi}
{
\frac{N_c^2}{(N_c^2-1)^2}
} 3(2 C_A - C_F) 
\Bigg\{ -\frac{(N_c^2-1)}{2N_c^2} \frac{(t^2+u^2)}{tu} \hbox{ln}\left(\frac{p_T^2}{s}\right) \\
{} &&
-\frac{(N_c^2-1)}{2}\left[\frac{u^2-t^2}{tu} + \frac{2(u-t)}{s} \right]\hbox{ln} \left(\frac{u}{t}\right)
\Bigg\}\left[\frac{\hbox{ln}^2(s_4/p_T^2)}{s_4}\right]_{+}
\, . 
\label{eq:gg2qqb2l}
\end{eqnarray}
A complete treatment of the kinematics and phase space integration can be
found in \cite{Beenakker:1988bq} and the plus-distributions are defined as in~\cite{Laenen:1998qw}. 
We note that the relative contribution of the above subprocess $gg \to q\bar{q}$ 
to the total cross section is numerically very small for both Tevatron and LHC energies, 
hence the differences observed in Eq.~(\ref{eq:gg2qqb2l}) are numerically
small in any application for collider phenomenology.

Let us now present the transverse momentum distributions of the inclusive jet
at both Tevatron ($\sqrt{S}=1.96$ TeV) and LHC ($\sqrt{S} = 7$ TeV).
Since we are interested in the perturbative convergence of the coefficient functions, 
we convolute these functions with just a set of PDFs extracted to a certain order.  
In our analysis, we use 
CTEQ6.6 ($\alpha_s(M_Z^2)=0.118$)~\cite{Nadolsky:2008zw} and 
ABM11 NNLO ($\alpha_s(M_Z^2)=0.1134$) \cite{Alekhin:2012ig} PDFs.
The strong coupling $\alpha_s$ is provided by the respective PDF sets through
{\tt LHAPDF} interface~\cite{Whalley:2005nh}. 
Throughout our analysis, we use the scale choice $\mu_F = \mu_R = p_T$, 
where $p_T$ is the transverse momentum of the observed jet. 
We present our distributions for jet transverse momentum in the central rapidity region
$0 \leq |y| \leq 0.5$ for LHC and $0 \leq y \leq 0.4$ for Tevatron, 
where the parton fluxes are dominated by parton momentum fractions $x_1$ and $x_2$ 
of similar order, $y$ being the jet rapidity. 
Further, in the rest of the paper we use the following $K$-factors defined as:
\begin{eqnarray}
K^{{(1)}}  =  1 + \frac{\sigma^{(1)}}{\sigma^{(0)}} ,  && \quad 
K^{{(2)}}  =  1 + \frac{\sigma^{(2)}}{\sigma^{(0)}} , \label{k1k2}\\
\nonumber {} && {} \\
K^{{(NLO)}}  = 1 + \frac{\sigma^{(NLO)}} {\sigma^{(0)}} , && \quad
K^{(NNLO*)}  =  1 + \frac{\sigma^{(NLO)} + \sigma^{(2)}}{\sigma^{(0)}} \label{knlo}
\, ,
\end{eqnarray}
where $\sigma^{(0)}$ is the LO cross section, $\sigma^{(1)}$ 
and $\sigma^{(2)}$ are respectively the $1$-loop and $2$-loop threshold corrections
expanded to only NLL accuracy and $\sigma^{(NLO)}$ is the exact NLO correction
to the cross section.

\begin{figure}[hhh]
\center            
  \includegraphics[width=0.46\textwidth]{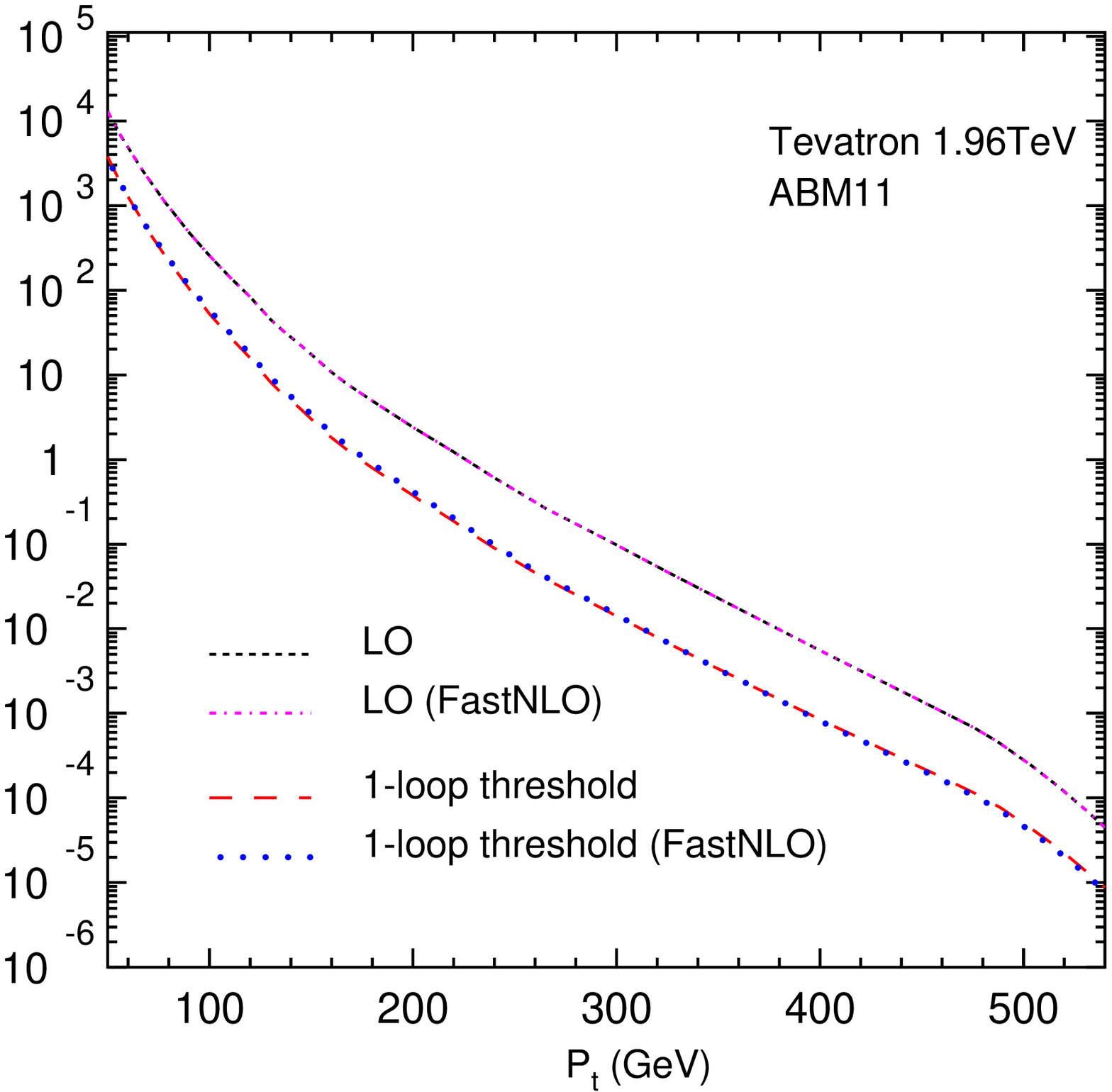}
  \includegraphics[width=0.46\textwidth]{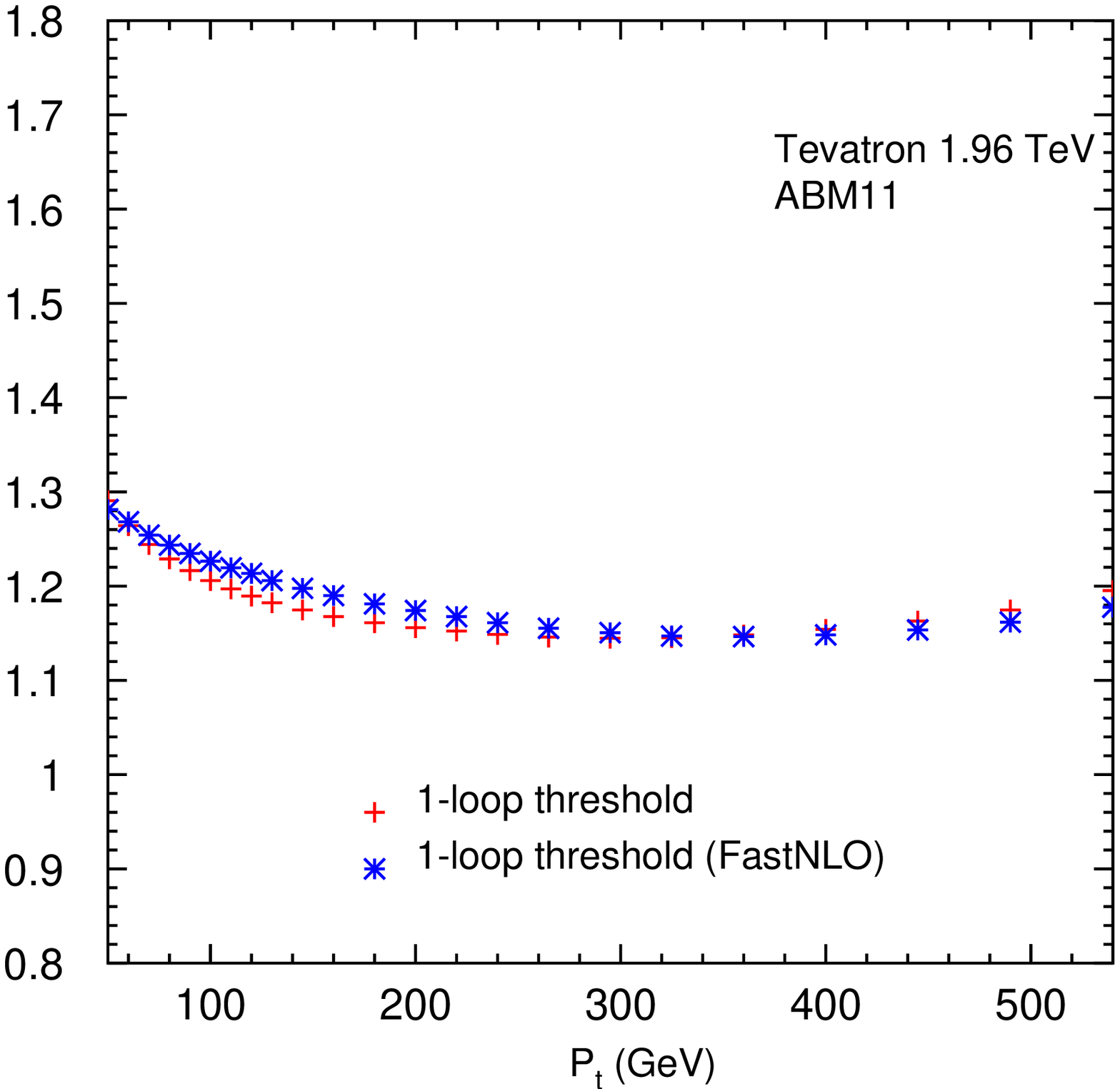}
\setlength{\unitlength}{1cm}                               
\caption{\label{tev-1loop} 
LO results and 1-loop threshold corrections $\sigma^{(1)}$ for the transverse momentum distribution 
of the jet (left) and the corresponding $K$-factor $K^{(1)}$ (right) at Tevatron.
} 
\vskip 1.0cm
\center            
  \includegraphics[width=0.46\textwidth]{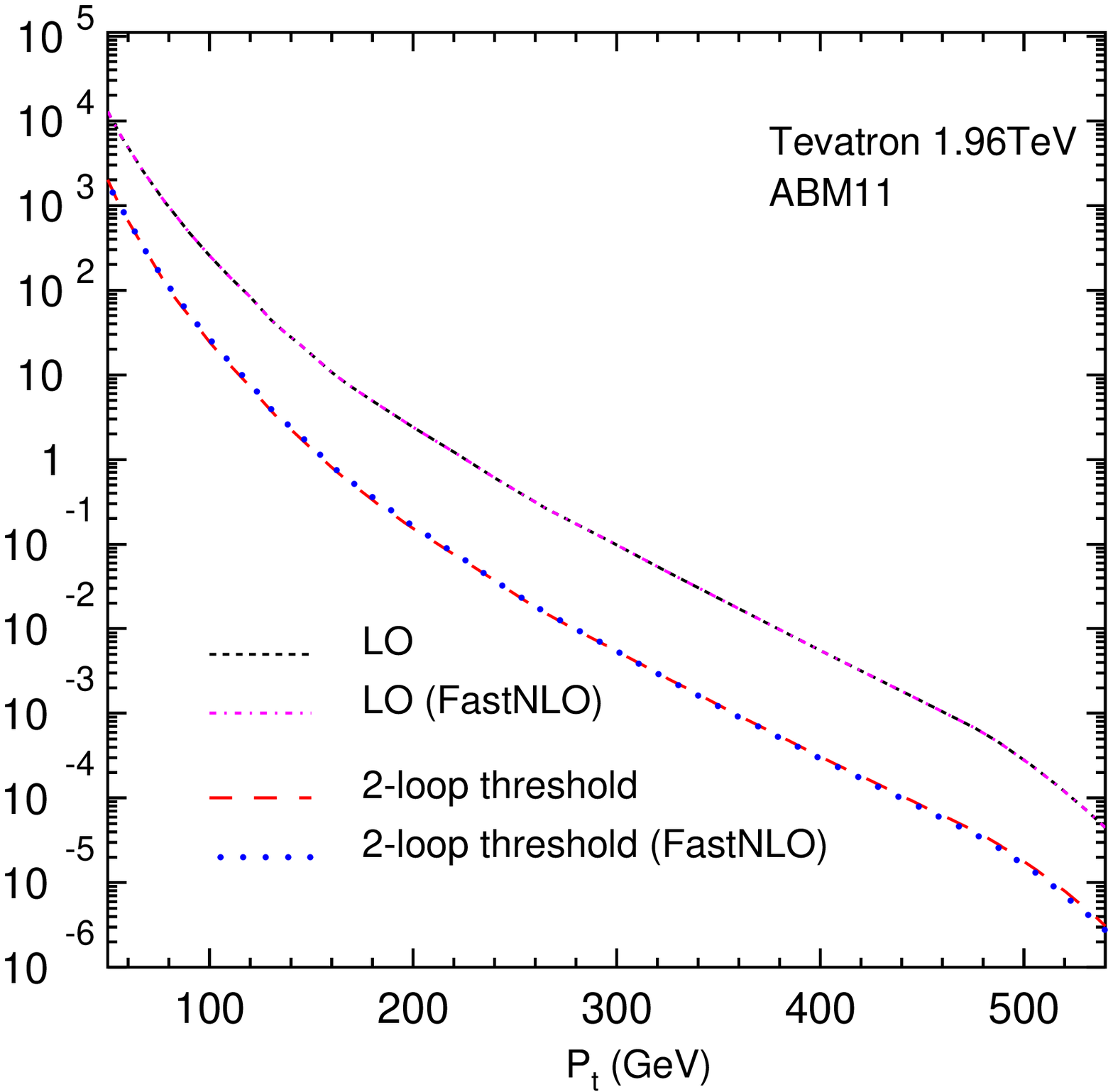}
  \includegraphics[width=0.46\textwidth]{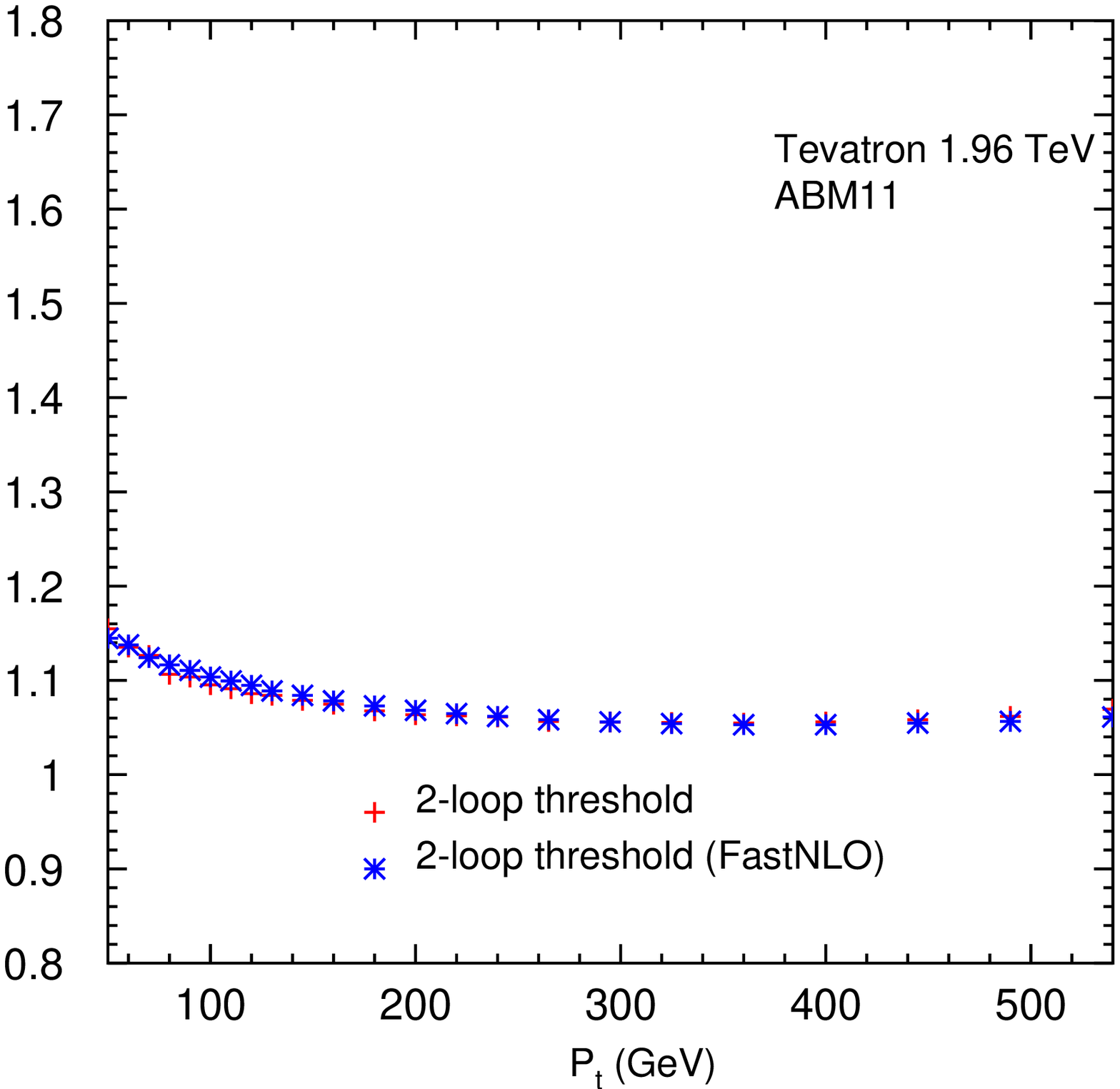}
\setlength{\unitlength}{1cm}
\caption{\label{tev-2loop} 
LO results and 2-loop threshold corrections $\sigma^{(2)}$ for the transverse momentum distribution 
of the jet (left) and the corresponding $K$-factor $K^{(2)}$ (right) at Tevatron.
} 
\end{figure}

\begin{figure}[hhh]
\center            
  \includegraphics[width=0.46\textwidth]{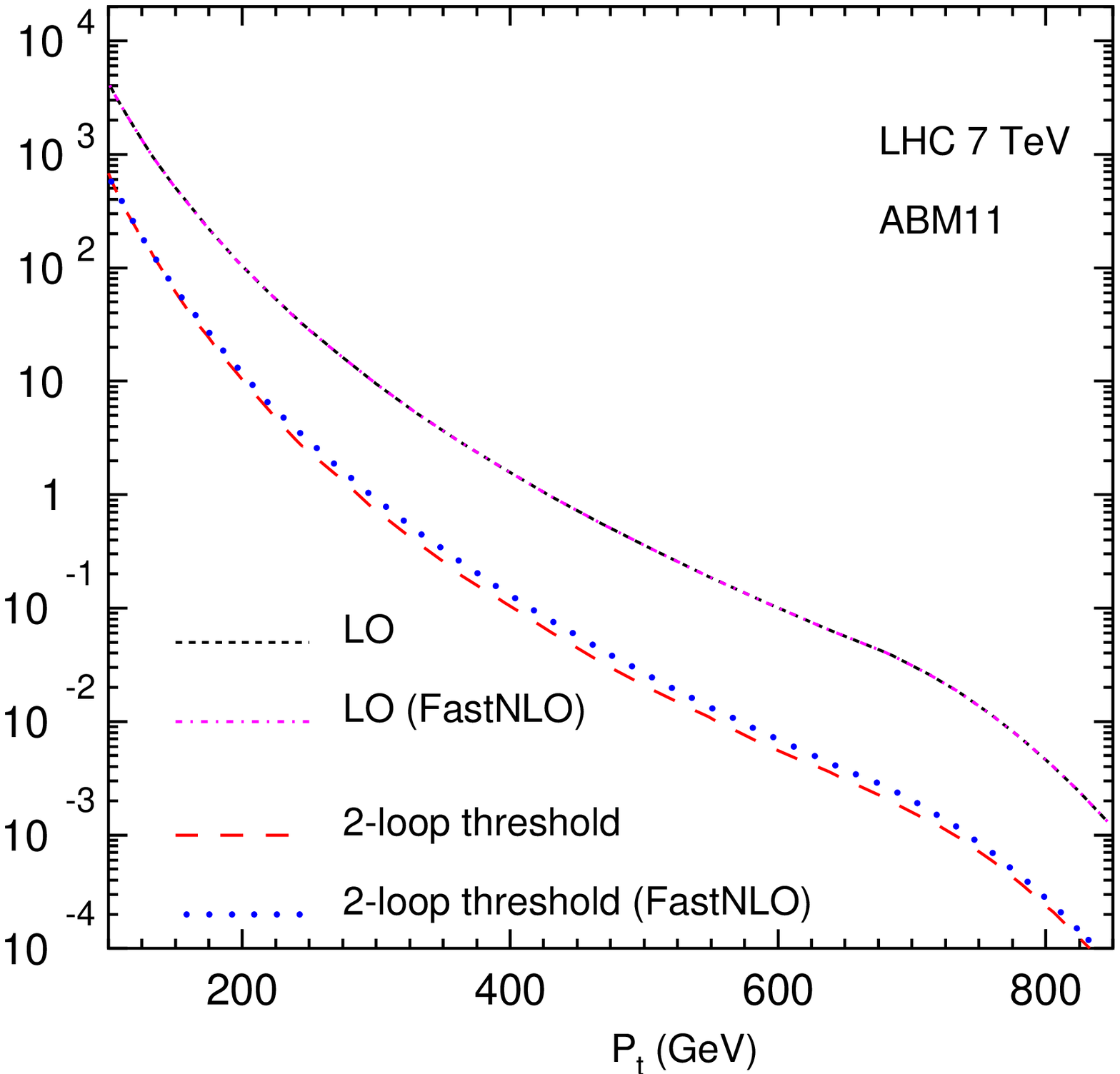}
  \includegraphics[width=0.46\textwidth]{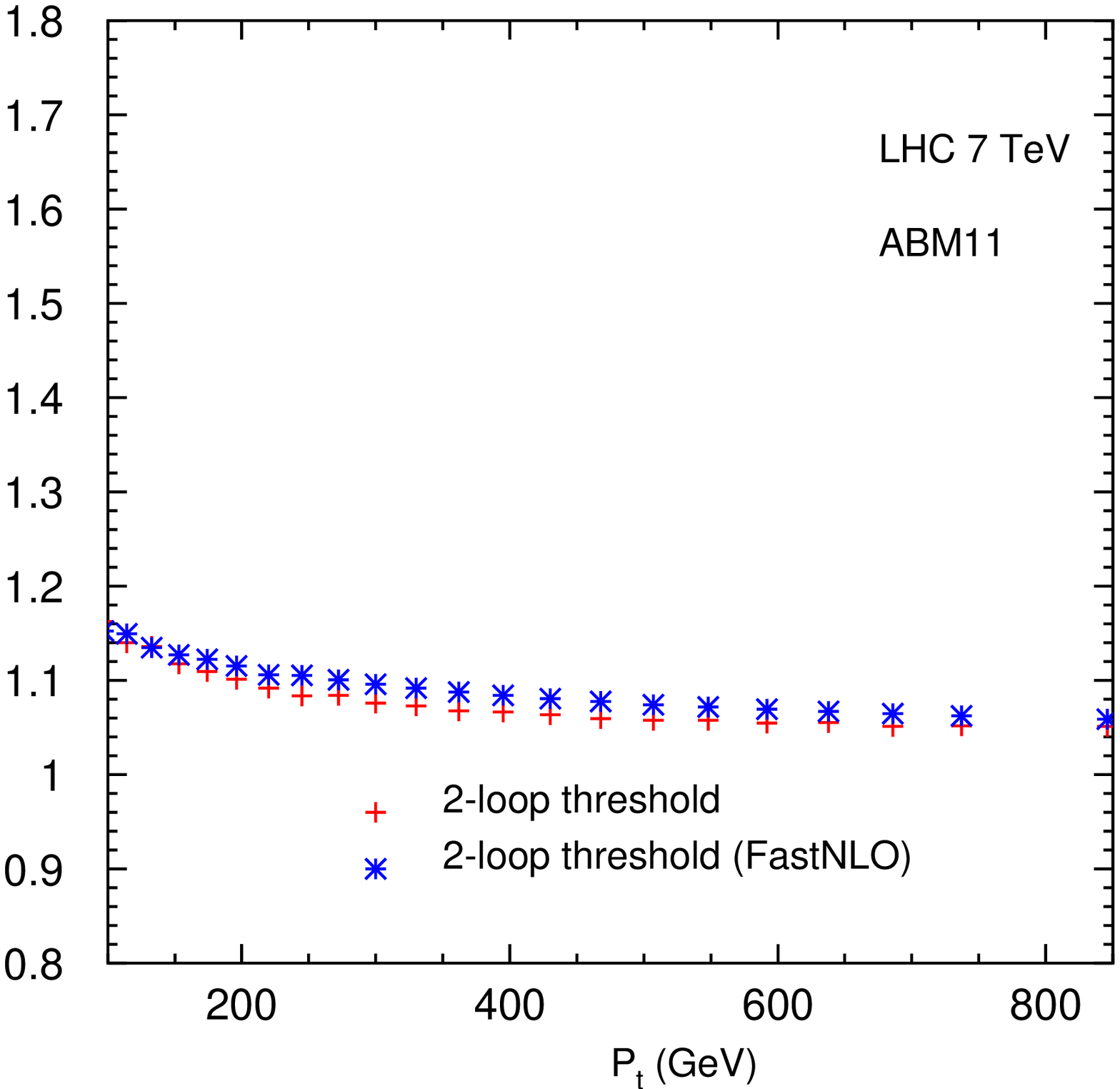}
\setlength{\unitlength}{1cm}                               
\caption{\label{lhc-2loop}
Same as Fig.~\ref{tev-2loop} for the $\sqrt{S}=7$ TeV LHC.}
\vskip 1.0cm
\center            
  \includegraphics[width=0.46\textwidth]{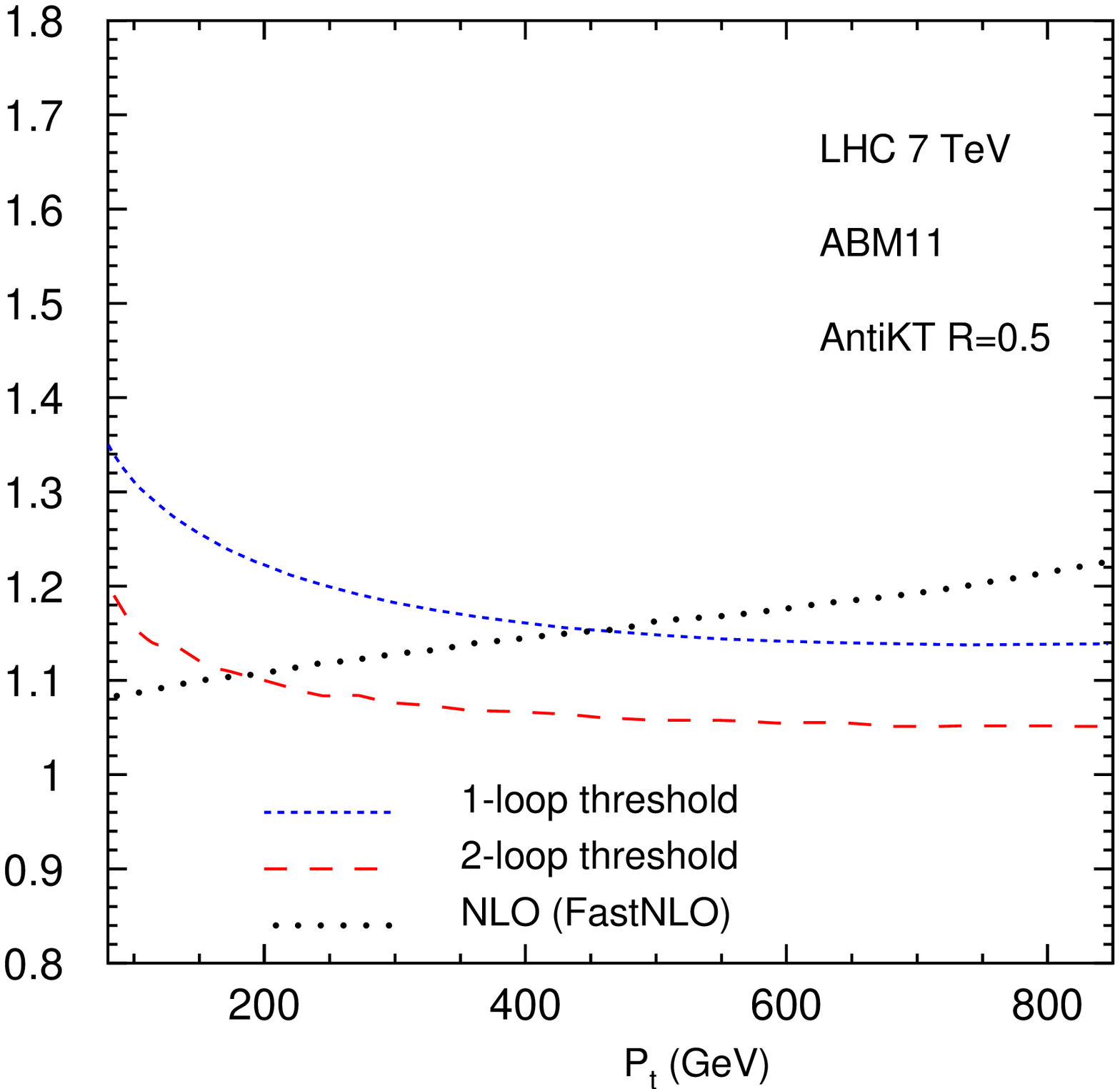}
  \includegraphics[width=0.46\textwidth]{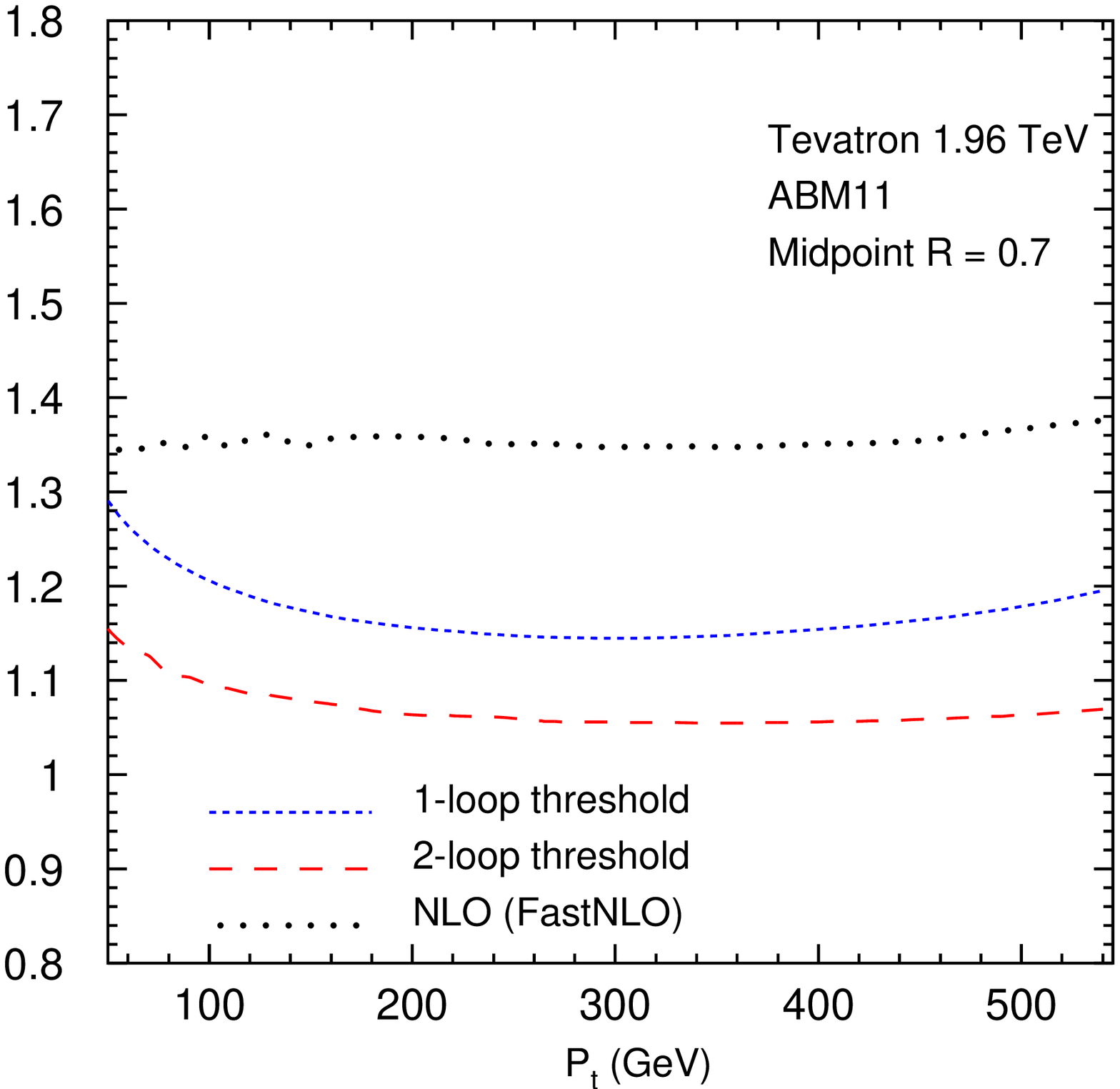}
\setlength{\unitlength}{1cm}                               
\caption{\label{nlo-threshold} 
$K$-factors $K^{(1)}$, $K^{(2)}$ and $K^{(NLO)}$ 
defined with respect to 1-loop threshold corrections, 2-loop threshold
corrections and the exact NLO results for $\sqrt{S}=7$ TeV LHC (left) and for Tevatron (right).
} 
\end{figure}

As a first check, we compare our numerical results with those obtained from {\tt FastNLO}~\cite{Wobisch:2011ij,Britzger:2012bs}. 
In the left panel of Fig.~\ref{tev-1loop}, we show the comparison 
of LO cross sections and $1$-loop threshold corrections $\sigma^{(1)}$ 
for Tevatron at $\sqrt{S}=1.96$ TeV center-of-mass (cms) energy 
and in the right panel of Fig.~\ref{tev-1loop} 
the corresponding $K$-factor $K^{(1)}$ as defined in Eq.~(\ref{k1k2}).  
Similar plots for $2$-loop threshold corrections $\sigma^{(2)}$ and the $K$-factors $K^{(2)}$ 
are presented in Fig.~\ref{tev-2loop} for the Tevatron at $\sqrt{S}=1.96$ TeV 
and in Fig.~\ref{lhc-2loop} for $\sqrt{S}=7$ TeV LHC.
In all cases, we find that our results are well 
in agreement with those obtained from {\tt FastNLO}.
For the $2$-loop threshold corrections $\sigma^{(2)}$ this constitutes an
independent check of~\cite{Kidonakis:2000gi} and confirms that possible 
differences in the analytical expressions, cf. Eq.~(\ref{eq:gg2qqb2l}), have
small numerical impact.

Next, we validate the threshold corrections by comparing 
them with the fixed order NLO results in the perturbation theory.  
In Fig.~\ref{nlo-threshold}, we present the $K$-factors $K^{(1)}$, $K^{(2)}$ and $K^{(NLO)}$. 
The NLO results for $K^{(NLO)}$ are read from the grids of {\tt FastNLO}.
In the case of LHC at $\sqrt{S}=7$ TeV cms (left panel in Fig.~\ref{nlo-threshold})
these are used in the CMS inclusive jet data analysis \cite{CMS:2011ab}
together with the anti-$k_t$ jet algorithm~\cite{Cacciari:2008gp} with $R=0.5$. 

We observe in Fig.~\ref{nlo-threshold} that $K^{(1)}$ and $K^{(2)}$ are sizable, 
of the order ${\cal O}(1.1)$ to ${\cal O}(1.2)$ at large $p_T$. 
The high $p_T$ region of the jet corresponds to the threshold region $s_4 =0$, 
where the phase space for the gluon radiation is limited.  
In this region, in particular the $1$-loop threshold corrections are expected 
to reproduce the exact fixed order NLO QCD corrections, i.e., $K^{(1)} \simeq K^{(NLO)}$, 
as a result of the dominance of the Sudakov logarithms in the perturbation expansion.
However, as can be seen from Fig.~\ref{nlo-threshold}, this is not quite the case.
Far away, from the threshold region, at small $p_T$, the threshold corrections in $K^{(1)}$ 
are found to be larger than $K^{(NLO)}$ for $p_T < 400$ GeV and 
for lower $p_T$ values (for about $p_T < 200$ GeV), even $K^{(2)}$ is found to exceed $K^{{(NLO)}}$. 
This indicates, that the $2$-loop threshold corrections, as such, 
in this region of phase space are subject to very large theory uncertainties and 
cannot be used in the relevant experimental data analysis.

\begin{figure}[hhh]
\center            
  \includegraphics[width=0.75\textwidth]{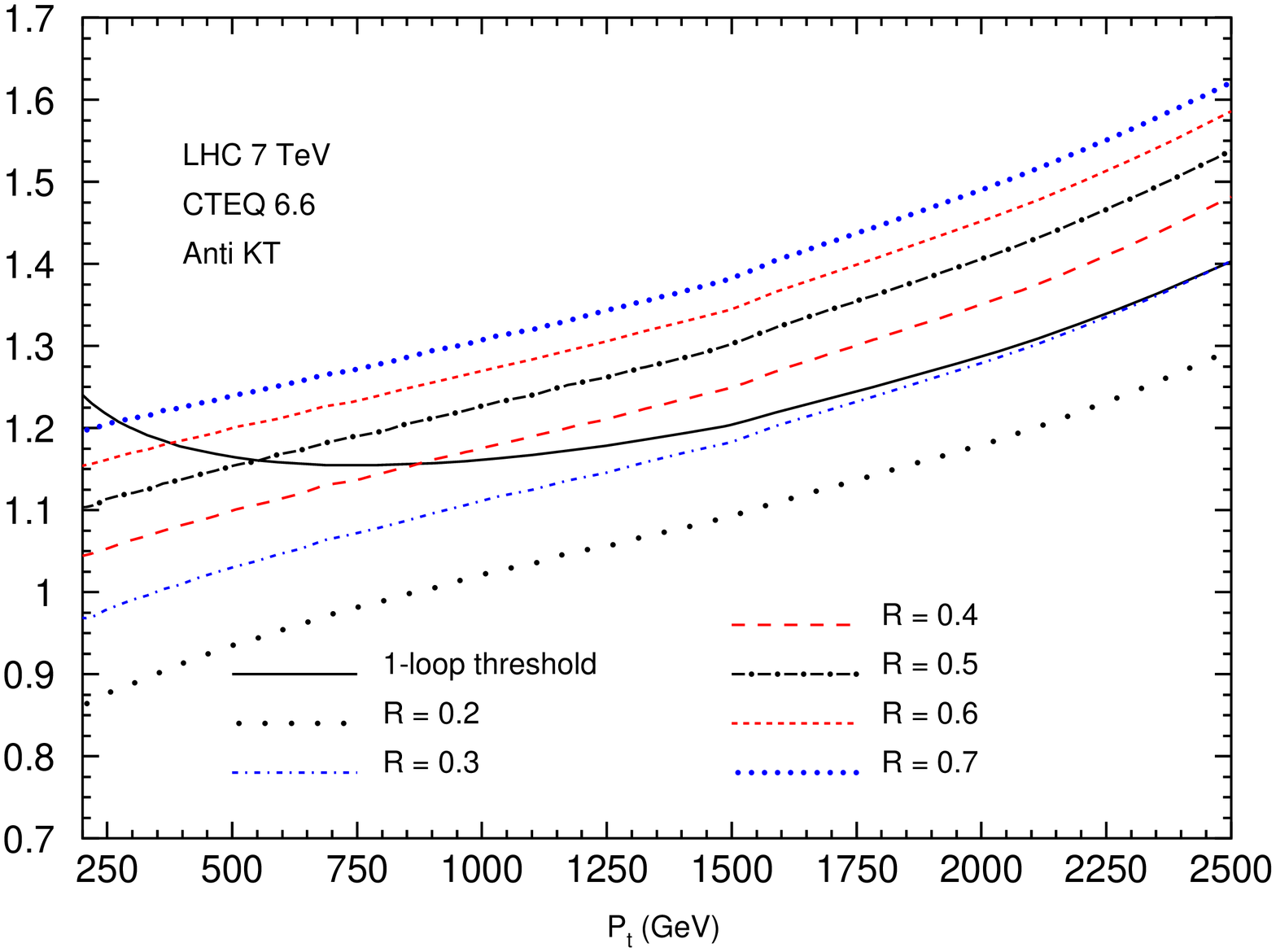}
\setlength{\unitlength}{1.0cm}
\caption{\label{lhc-cone-variation}
NLO $K$-factors $K^{(NLO)}$ for inclusive jet production as a function of the parameter $R$ in the anti-$k_t$ jet 
algorithm, computed for $\sqrt{S}=7$ TeV LHC. The solid line corresponds to the one-loop threshold 
corrections $K^{(1)}$ at NLL accuracy.
}
\center            
  \includegraphics[width=0.75\textwidth]{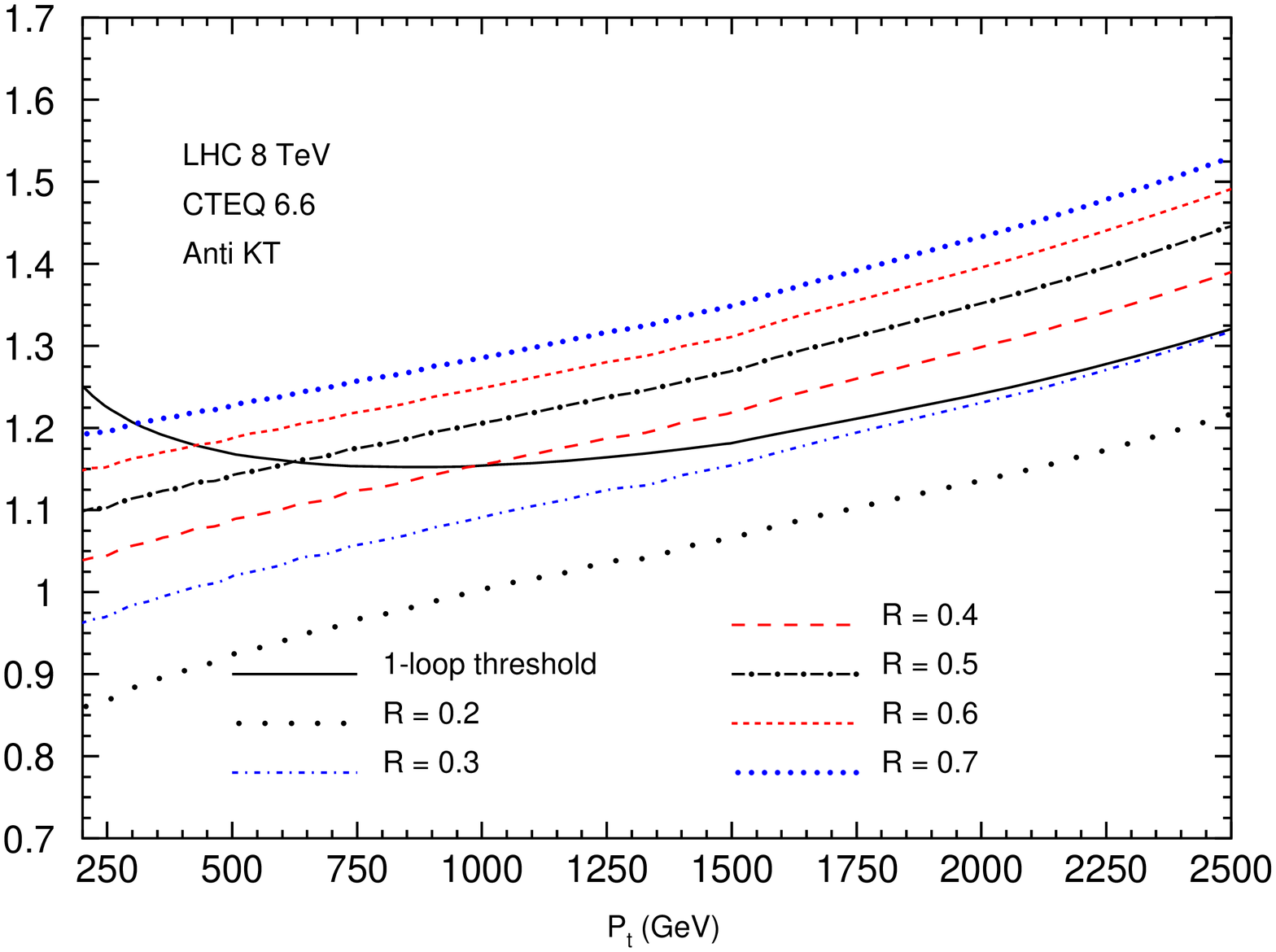}
\setlength{\unitlength}{1.0cm}
\caption{\label{lhc-cone-variation-8tev}
Same as Fig.~\ref{lhc-cone-variation} for the $\sqrt{S}=8$ TeV LHC.
}
\end{figure}

\begin{figure}[hhh]
\center            
  \includegraphics[width=0.75\textwidth]{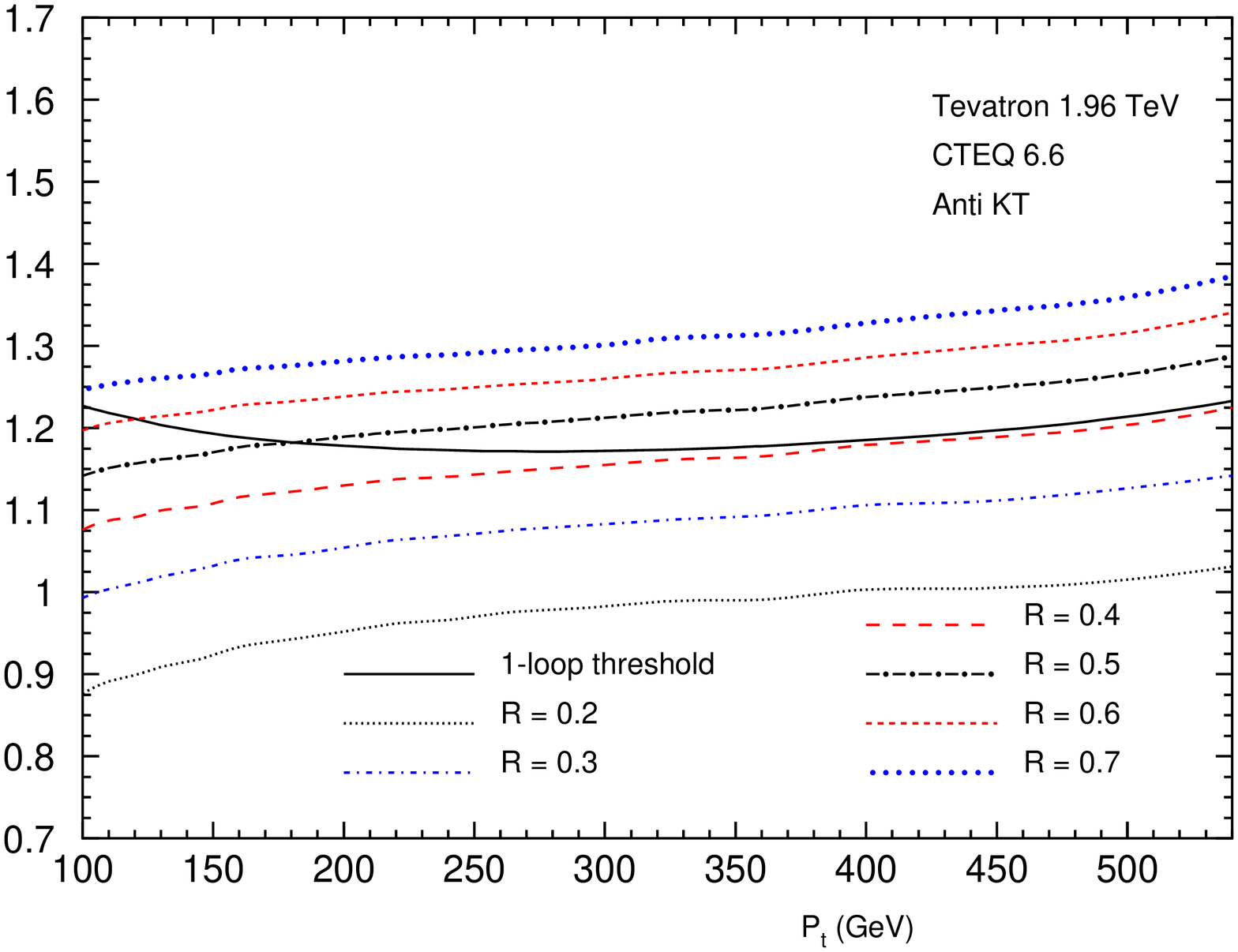}
\setlength{\unitlength}{1.0cm}
\caption{\label{tev-cone-variation}
Same as Fig.~\ref{lhc-cone-variation} for the Tevatron.
} 
\center            
  \includegraphics[width=0.75\textwidth]{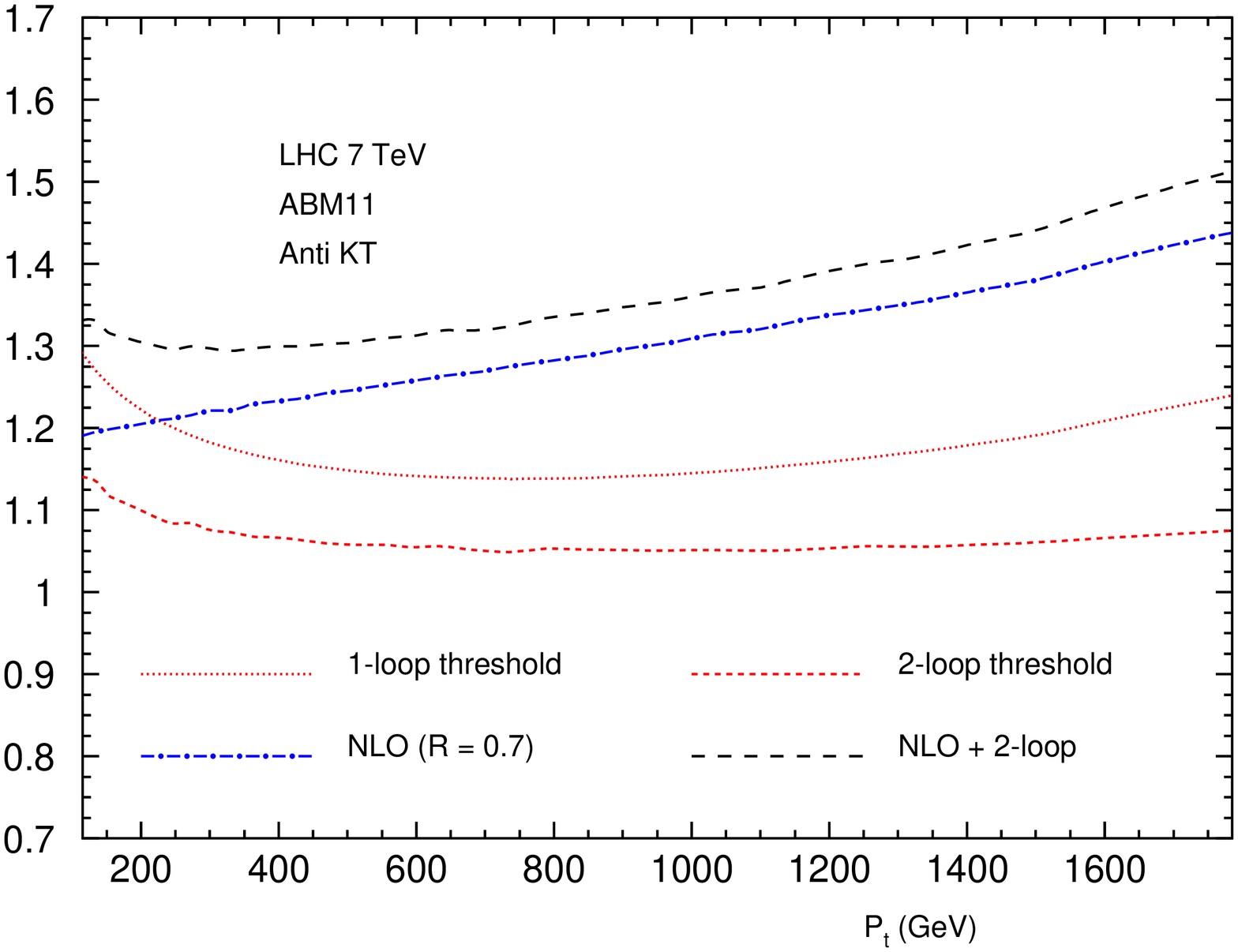}
\setlength{\unitlength}{1.0cm}
\caption{\label{lhc.7tev.kf}
Comparison of $K$-factors $K^{(1)}$, $K^{(2)}$, $K^{(NLO)}$ and $K^{(NNLO*)}$
for $1$-loop threshold, $2$-loop threshold, NLO and
NLO + $2$-loop (NNLO*) cross sections computed for $\sqrt{S}=7$ TeV LHC.
}
\end{figure}

In order to clarify the deviations between $K^{(1)}$ and $K^{(NLO)}$ 
illustrated in Fig.~\ref{nlo-threshold} we study the dependence on $R$. 
We compute the NLO cross sections as a function of $R$ for inclusive jet production at LHC and Tevatron. 
For this computation, we use {\tt NLOJET++} program, 
anti-$k_t$ jet algorithm \cite{Cacciari:2008gp} from {\tt FastJet}~\cite{Cacciari:2011ma}. 
and CTEQ6.6 PDFs~\cite{Nadolsky:2008zw}.
In Figs.~\ref{lhc-cone-variation} and \ref{lhc-cone-variation-8tev}
we present our results in terms of $K^{{(NLO)}}$ for $\sqrt{S}=7$ TeV and $8$ TeV LHC 
by varying $R$ from $0.2$ to $0.7$ and by considering $p_T$ of jet as high as $2500$ GeV.  
Likewise, Fig.~\ref{tev-cone-variation} displays the results for the Tevatron
Run II case using the anti-$k_t$ jet algorithm and varying $R$ from $0.2$ to $0.7$.
As can be seen from those figures, the NLO QCD cross sections increase with the cone size $R$. 
Further, $K^{{(NLO)}}$ is less than unity for smaller $p_T$ values and for smaller $R$ values, 
because the ${\cal O}(\alpha_s)$ QCD corrections are negative in this region.  
On the contrary for higher $R (>0.4)$ values, $K^{{(NLO)}}$ is always greater than unity. 
Moreover, the NLO QCD corrections do increase by about $30 \%$ as $R$ varies from $0.2$ to $0.7$, 
regardless of the value of $p_T$ in the range considered here.

\begin{figure}[hhh]
\center            
  \includegraphics[width=0.75\textwidth]{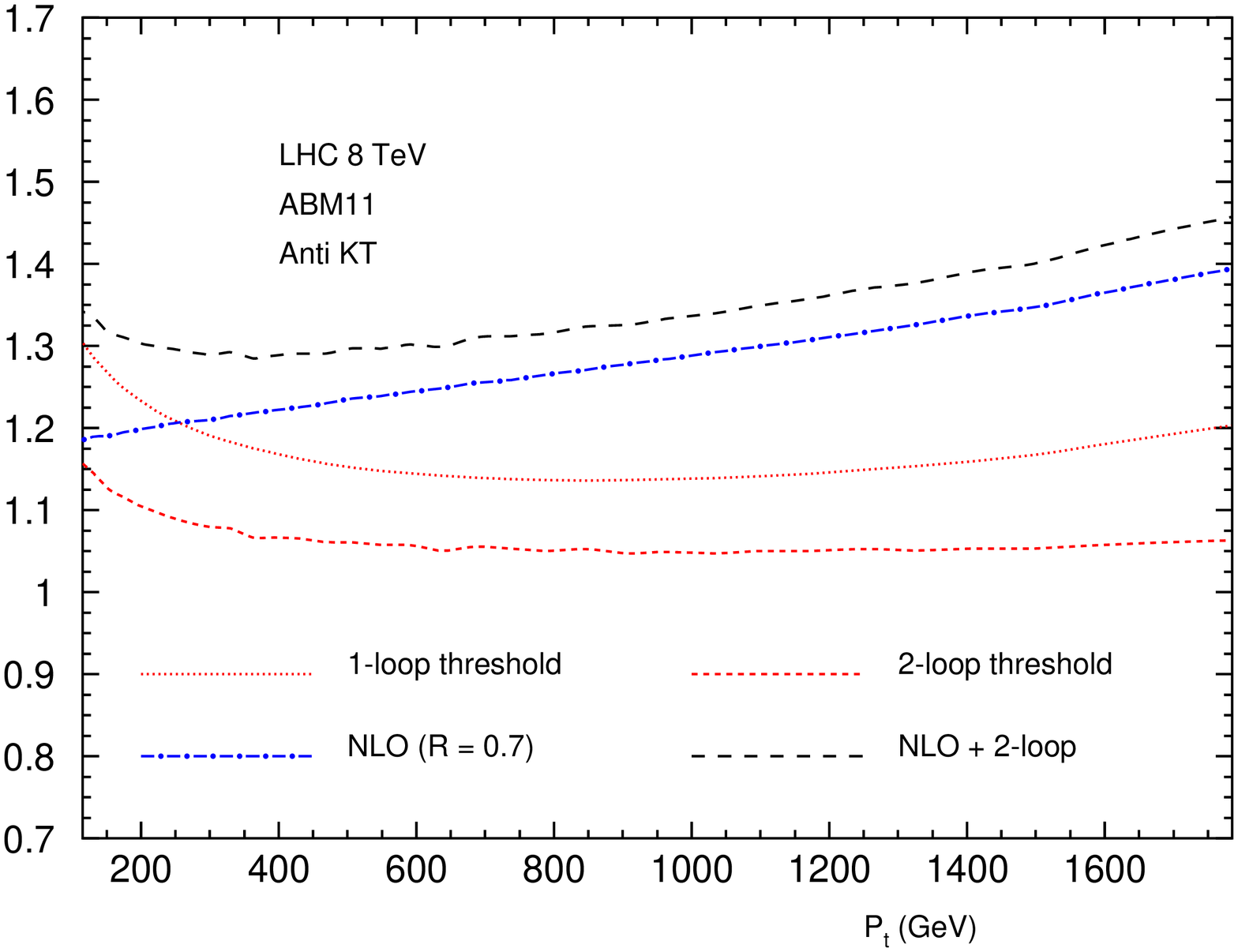}
\setlength{\unitlength}{1.0cm}
\caption{\label{lhc.8tev.kf}
Same as Fig.~\ref{lhc.7tev.kf} for the $\sqrt{S}=8$ TeV LHC.
}
\end{figure}

It is therefore quite revealing to compare these NLO corrections with the
$1$-loop threshold corrections as done in Figs.~\ref{lhc-cone-variation}-\ref{tev-cone-variation}.
There, in Fig.~\ref{lhc-cone-variation} for $\sqrt{S}=7$ TeV LHC, 
$K^{{(1)}}$ decreases with increasing $p_T$ up to about $800$ GeV and then increases with $p_T$. 
At very large $p_T$ the threshold logarithms are dominant and we observe for the $K$-factors 
$K^{{(1)}}$ and $K^{{(NLO)}}$ the same rising behavior in this region.
Interestingly, in the high $p_T$ region the approximation 
which is independent of $R$ coincides with the 
exact NLO result only when the latter is computed for smaller $R$ values of about $0.3$,
i.e., $K^{{(1)}} \simeq K^{{(NLO)}}$ for $R=0.3$ for the LHC, cf. Figs.~\ref{lhc-cone-variation} and \ref{lhc-cone-variation-8tev}.
Likewise, for the Tevatron the $1$-loop threshold corrections 
are comparable to the exact NLO ones for the cone size of about $R=0.4$ in the high $p_T$ region, 
cf. Fig.~\ref{tev-cone-variation}.
In Figs.~\ref{lhc.7tev.kf} and \ref{lhc.8tev.kf}, we present the $K$-factors
$K^{{(1)}}$, $K^{{(2)}}$, $K^{{(NLO)}}$ and $K^{{(NNLO*)}}$  for $\sqrt{S}=7$ TeV and 8 TeV LHC respectively
for a cone size of $R=0.7$.
In summary, the absence of any dependence on the jet's cone size $R$ in the threshold corrections 
implies a very large theoretical uncertainty inherent in~\cite{Kidonakis:2000gi}.

In discussing our findings, it is worth noting here that the corresponding $2$-loop threshold corrections
for the Tevatron illustrated in Figs.~\ref{tev-2loop} and \ref{tev-cone-variation} have been used in the determination
of the strong coupling constant from the Tevatron inclusive jet cross section data \cite{Abazov:2009nc}
by considering the jet transverse momentum in the range $50 < p_T < 145$ GeV. The corresponding 
theory predictions are obtained from MSTW 2008 PDF sets. 
In this analysis, the strong coupling constant 
obtained from pure NLO perturbative QCD corrections is determined to be $\alpha_s(M_Z^2) = 0.1201$ 
while the inclusion of the $2$-loop threshold corrections has decreased its
central value to $\alpha_s (M_Z^2) = 0.1161$.

Moreover, another remark to be made in the discussion of Figs.~\ref{lhc-cone-variation} and \ref{lhc-cone-variation-8tev} 
is that the 1-loop threshold corrections 
in the low $p_T$ region of the jet ($p_T < 500$ GeV), 
are much higher than the exact NLO QCD corrections computed for all values of $R < 0.7$.
For improved approximations beyond NLL, it is required to systematically
include also the hard matching functions $H^{(1)}$ that can be extracted from the finite parts of the virtual corrections 
in the NLO computation. 
Such an analysis, but using different kinematics, has been done in~\cite{deFlorian:2005yj} 
wherein the logarithms of the kind $\alpha_s^{k}\hbox{ln}^{2k}(1-x_T^2)$ are 
resummed at NLL accuracy.  
An extension to this work has also been done in \cite{deFlorian:2007fv}
where the integration is done over jet mass defined in terms of the cone size $R$.
However, for the present case using $s_4$ kinematics where the logarithms of type 
$[\ln^{l}(s_4/p_T^2)/s_4]_+$ are considered, the hard matching functions 
are expected to be small in the threshold region as they are independent of 
threshold logarithms and the relevant parton fluxes in this region fall rapidly.

Further necessary improvements thus concern the extension of the threshold corrections to NNLL accuracy,
a proper treatment of the jet's kinematics and cone size and, of course, 
the completion of the exact NNLO QCD corrections~\cite{Ridder:2013mf}.
Unrelated, though also necessary is inclusion of the electro-weak corrections at NLO to hadro-production of jets
possibly the effect of electro-weak Sudakov logarithms, see, e.g.,~\cite{Dittmaier:2013hha,Kuhn:2001hz}.

To summarize, we have computed the threshold corrections to inclusive jet 
production at hadron colliders in the soft-gluon resummation formalism. 
We find that that our results are in agreement with those in the literature
apart from few typographical errors. Furthermore, we have investigated the phenomenology 
of these threshold corrections by comparing them expanded to $1$-loop level at NLL 
accuracy with the exact NLO results. We have also studied the dependence of the 
exact NLO results on the cone size $R$. These QCD threshold corrections are 
better comparable in the high $p_T$ region with the exact NLO QCD corrections 
only when the latter are computed for smaller cone sizes, about $R=0.3$ and $R=0.4$ 
for LHC and Tevatron.  For the LHC at $\sqrt{S}=7$ TeV cms energy, our analysis indicates 
that applying these threshold corrections for $p_T < 500$ GeV can lead to large 
uncertainties and in particular potential theoretical uncertainties for $p_T < 200$ GeV.  
On the contrary, for higher $p_T$ values near threshold region, they underestimate
the fixed order results in the perturbation theory for typical values of R used 
in jet analysis at LHC experiments.

\subsection*{Acknowledgments}
We are thankful to D.~Britzger, K.~Rabbertz and M.~Wobisch for useful discussions on {\tt FastNLO} 
and for providing us with a {\tt FastNLO} table for $1$-loop threshold corrections at Tevatron. 
We also thank N.~Kidonakis for his comments on the analytical part of this computation.

This work has been supported by Deutsche Forschungsgemeinschaft in
Sonderforschungs\-be\-reich SFB 676 
and by the European Commission through contract PITN-GA-2010-264564 ({\it LHCPhenoNet}).

{\footnotesize


}

\newpage

\clearpage

\clearpage

\clearpage

\end{document}